\documentclass[twocolumn,superscriptaddress]{revtex4}
\usepackage{amssymb}
\usepackage{amsmath}

\usepackage{graphicx}
\usepackage{dcolumn}
\usepackage{bm}

\newfont{\Fr}{eufm10}

\begin{document}

\title{Dark torsion as the cosmic speed-up}
\author{Gabriel R. Bengochea}
\email{gabriel@iafe.uba.ar}
\thanks{CONICET Fellow}
\affiliation{Instituto de Astronom\'\i a y F\'\i sica del Espacio, Casilla de Correo 67,
Sucursal 28, 1428 Buenos Aires, Argentina}
\affiliation{Departamento de F\'\i sica, Facultad de Ciencias Exactas y Naturales,
Universidad de Buenos Aires, Ciudad Universitaria, Pabell\'on I, 1428 Buenos
Aires, Argentina}
\author{Rafael Ferraro}
\email{ferraro@iafe.uba.ar}
\thanks{Member of Carrera del Investigador Cient\'{\i}fico (CONICET,
Argentina)}
\affiliation{Instituto de Astronom\'\i a y F\'\i sica del Espacio, Casilla de Correo 67,
Sucursal 28, 1428 Buenos Aires, Argentina}
\affiliation{Departamento de F\'\i sica, Facultad de Ciencias Exactas y Naturales,
Universidad de Buenos Aires, Ciudad Universitaria, Pabell\'on I, 1428 Buenos
Aires, Argentina}

\begin{abstract}
It is shown that the recently detected acceleration of the
universe can be understood by considering a modification of the
teleparallel equivalent of General Relativity (TEGR), with no need
of dark energy. The solution also exhibits phases dominated by
matter and radiation as expected in the standard cosmological
evolution. We perform a joint analysis with measurements of the
most recent type Ia supernovae (SNe Ia), Baryon Acoustic
Oscillation (BAO) peak and estimates of the CMB shift parameter
data to constraint the only new parameter this theory has.
\end{abstract}

\pacs{Valid PACS appear here}
\maketitle


\section{Introduction}

The discovery of an unexpected diminution in the observed energy
fluxes coming from type Ia supernovae \cite{perlmutter-otros,
kowalski} has opened one of the most puzzling and deepest problems
in cosmology today. These observations have been interpreted as
solid evidence for an accelerating universe dominated by something
called dark energy. Although the cosmological constant seems to be
the simplest explanation for the phenomenon, several dynamical
scenarios have been tried out since 1998 (see e.g., \cite{sahni,
padma, frieman}). While some authors sustain the idea of the
existence of a dark energy, others propose modifications of the
Einstein-Hilbert Lagrangian known as $f(R)$ (\cite{buch, staro,
ker, barrow1, barrow2, carroll, starobinsky, hu, odintsov1} or
\cite{odintsov2, capo, sotiriou} for recent reviews) as a way to
obtain a late accelerating expansion. A great difficulty these
theories have, from the point of view of the metric formalism, is
that the resulting field equations are 4th order equations, which
in many cases makes
these hard to analyze. Besides, the simplest cases of the kind $%
f(R)=R-\beta/R^n$ have shown difficulties with, weak field tests \cite%
{chiba, olmo}, gravitational instabilities \cite{dolgov} and do
not present a matter dominated era previous to the acceleration
era \cite{amendola1, amendola2}. Alternatively, the Palatini
variational approach for such $f(R)$ theories leads to 2nd order
field equations, and some authors have achieved to put
observational constraints to these theories \cite{amar, fay,
santos}. However in many cases the equations are still hard to
work with, as evidenced by the functional form of the modified
Friedmann equation for a generic $f(R)$. Recently, models based on
modified teleparallel gravity were presented as an alternative to
inflationary models \cite{franco,franco2}. In this paper we show a
cosmological solution for the acceleration of the universe by
means of a sort of theories of modified gravity, namely $f(L_T)$,
based on a modification of the teleparallel equivalent of General
Relativity (TEGR) Lagrangian \cite{einstein, hayashi} where the
torsion will be the responsible of the observed acceleration of
the universe, and the field equations will always be 2nd order
equations.

\section{General considerations. Field equations}

Teleparallelism \cite{einstein, hayashi} uses as dynamical object a vierbein
field ${\mathbf{e}_i(x^\mu)}$, $i=0, 1, 2, 3$, which is an orthonormal basis
for the tangent space at each point $x^\mu$ of the manifold: $\mathbf{e}%
_i\cdot\mathbf{e}_j=\eta_{i\, j}$, where $\eta_{i\, j}=diag (1,-1,-1,-1)$.
Each vector $\mathbf{e}_i$ can be described by its components $e_i^\mu$, $%
\mu=0, 1, 2, 3$ in a coordinate basis; i.e. $\mathbf{e}_i=e^\mu_i\partial_\mu
$. Notice that latin indexes refer to the tangent space, while greek indexes
label coordinates on the manifold. The metric tensor is obtained from the
dual vierbein as $g_{\mu\nu}(x)=\eta_{i\, j}\, e^i_\mu (x)\, e^j_\nu (x)$.
Differing from General Relativity, which uses the torsionless Levi-Civita
connection, Teleparallelism uses the curvatureless Weitzenb\"{o}ck
connection \cite{weit}, whose non-null torsion is
\begin{equation}  \label{torsion2}
{T}^\lambda_{\:\mu\nu}=\overset{\mathbf{w}}{\Gamma}^\lambda_{\nu\mu}-\overset%
{\mathbf{w}}{\Gamma}^\lambda_{\mu\nu}=e^\lambda_i\:(\partial_\mu
e^i_\nu-\partial_\nu e^i_\mu)
\end{equation}
This tensor encompasses all the information about the gravitational field.
The TEGR Lagrangian is built with the torsion (\ref{torsion2}), and its
dynamical equations for the vierbein imply the Einstein equations for the
metric. The teleparallel Lagrangian is \cite{hayashi,maluf,arcos},
\begin{equation}  \label{lagTele}
L_T=S_\rho^{\:\:\:\mu\nu}\:T^\rho_{\:\:\:\mu\nu}
\end{equation}
where:
\begin{equation}  \label{S}
S_\rho^{\:\:\:\mu\nu}=\frac{1}{2}\Big(K^{\mu\nu}_{\:\:\:\:\rho}+\delta^\mu_%
\rho \:T^{\theta\nu}_{\:\:\:\:\theta}-\delta^\nu_\rho\:
T^{\theta\mu}_{\:\:\:\:\theta}\Big)
\end{equation}
and $K^{\mu\nu}_{\:\:\:\:\rho}$ is the contorsion tensor:
\begin{equation}  \label{K}
K^{\mu\nu}_{\:\:\:\:\rho}=-\frac{1}{2}\Big(T^{\mu\nu}_{\:\:\:\:\rho}
-T^{\nu\mu}_{\:\:\:\:\rho}-T_{\rho}^{\:\:\:\:\mu\nu}\Big)
\end{equation}
which equals the difference between Weitzenb\"{o}ck and Levi-Civita
connections.

In this work the gravitational field will be driven by a Lagrangian density
that is a function of $L_T$. Thus the action reads
\begin{equation}  \label{accionTP}
I= \frac{1}{16\, \pi\, G}\, \int d^4x\:e\:f(L_T)
\end{equation}
where $e=det(e^i_\mu)=\sqrt{-g}$. The case $f(L_T)=L_T$ corresponds to TEGR.
If matter couples to the metric in the standard form then the variation of
the action with respect to the vierbein leads to the equations
\begin{eqnarray}
&&e^{-1}\partial_\mu(e\:S_i^{\:\:\:\mu\nu})f^{\prime}(L_T)-e_i^{\:\lambda}
\:T^\rho_{\:\:\:\mu\lambda}\:S_\rho^{\:\:\:\nu\mu}f^{\prime}(L_T)+  \nonumber
\\
&&+S_i^{\:\:\:\mu\nu}\partial_\mu(L_T)f^{\prime\prime}(L_T)+\frac{1}{4}%
\:e_i^\nu \:f(L_T)=  \nonumber \\
&&=4\:\pi\:G\:e_i^{\:\:\:\rho}\:T_\rho^{\:\:\:\nu}  \label{ecsmovim}
\end{eqnarray}
where a prime denotes differentiation with respect to $L_T$,
$S_i^{\:\:\mu\nu}=e_i^{\:\:\rho}S_\rho^{\:\:\mu\nu}$ and
$T_{\mu\nu}$
is the matter energy-momentum tensor. The fact that equations (\ref{ecsmovim}%
) are 2nd order makes them simpler than the dynamical equations resulting in
$f(R)$ theories.

\section{Cosmological solution and observational constraints}

We will assume a flat homogeneous and isotropic FRW universe, so
\begin{equation}  \label{tetradasFRW}
e^i_\mu=diag(1,a(t),a(t),a(t))
\end{equation}
where $a(t)$ is the cosmological scale factor. By replacing in (\ref%
{torsion2}), (\ref{S}) and (\ref{K}) one obtains
\begin{equation}  \label{STFRW}
L_T=S^{\rho\mu\nu}T_{\rho\mu\nu}=-6\:\frac{\dot{a}^2}{a^2}=-6\:H^2
\end{equation}
$H$ being the Hubble parameter $H=\dot{a}\, a^{-1}$. As a remarkable
feature, the scale factor enters the invariant $L_T$ just through the Hubble
parameter. The substitution of the vierbein (\ref{tetradasFRW}) in (\ref%
{ecsmovim}) for $i=0=\nu$ yields
\begin{equation}  \label{FRWMod}
12\:H^2\:f^{\prime}(L_T)+f(L_T)=16\pi G\:\rho
\end{equation}
Besides, the equation $i=1=\nu$ is
\begin{equation}  \label{ec1-1}
48 H^2 f^{\prime\prime}(L_T) \dot{H}-f^{\prime}(L_T)[12 H^2 + 4 \dot{H}%
]-f(L_T)=16\pi G\:p
\end{equation}
In Eqs. (\ref{FRWMod}-\ref{ec1-1}), $\rho(t)$ and $p(t)$ are the total
density and pressure respectively. It can be easily derived that they
accomplish the conservation equation
\begin{equation}  \label{conservation}
\frac{d}{dt}(a^3\, \rho)\, =\, -3\, a^3\, H\, p
\end{equation}
whatever $f(L_T)$ is. Thus, if the state equation is $p=w\rho$ then $\rho$
evolves as $\rho\varpropto(1+z)^{3(1+w)}$ ($z$ is the cosmological redshift).

We are interested in obtaining an accelerated expansion without dark energy
but driven by torsion. For this we will try with a kind of $f(L_{T})$
theories:
\begin{equation}
f(L_{T})=L_{T}-\frac{\alpha }{(-L_{T})^{n}}  \label{fT}
\end{equation}%
being $\alpha $ and $n$ real constants to be determined by
observational constraints. Although the functional form of
(\ref{fT}) is similar to those considered in $f(R)$ literature,
now the guideline towards modified gravity is $H$ instead of $R$.
This fact gives to these theories another interesting feature
because $H$ is the most important cosmological variable. For later
times the term $-\alpha /(-L_{T})^{n}$ is dominant, while in early
times, when $H\rightarrow \infty $, General Relativity is
recovered. From (\ref{FRWMod}) along with (\ref{fT}), the modified
Friedmann equation results to be
\begin{equation}
H^{2}-\frac{(2n+1)\;\alpha }{6^{n+1}H^{2n}}=\frac{8}{3}\pi G\rho
\label{friedmodif}
\end{equation}%
(a functional dependence similar to the results other authors
arrived, through different theoretical motivations such as
\cite{dvali, goobar}).

Now, replacing $\rho =\rho _{mo}(1+z)^{3}+\rho _{ro}(1+z)^{4}$, and calling $%
\Omega _{i}=8\pi G\,\rho _{io}/(3H_{o}^{2})$ the contributions of
matter and radiation to the total energy density today, Eq.
(\ref{friedmodif}) becomes
\begin{equation}
y^{n}(y-B)\,=\,C  \label{friedmodif2}
\end{equation}%
where $y=H^{2}/H_{o}^{2}$, $B=\Omega _{m}(1+z)^{3}+\Omega
_{r}(1+z)^{4}$ and $C=\alpha (2n+1)\,(6H_{o}^{2})^{-(n+1)}$. The
evaluation of this equation for $z=0$ allows to rephrase the
constant $C$ as a function of $\Omega _{i}$ and $n$: $C=1-\Omega
_{m}-\Omega _{r}$. For $\alpha =0$ (then $1=\Omega _{m}+\Omega
_{r}$) the GR spatially flat Friedmann equation $H^{2}=H_{o}^{2}B
$ is retrieved. The case $n=0$ recovers the GR dynamics with
cosmological constant $\Omega _{\Lambda }=1-\Omega _{m}-\Omega
_{r}$. Notice the functional simplicity of (\ref{friedmodif})
compared with its analogue in $f(R)$ theories. Compared with GR,
$n$ is the sole new free parameter in (\ref{friedmodif2}), since
specifying the value of $n$ and $\Omega_m$ ($\Omega_r$) the value
of $\alpha$ (in units of $H^{2(n+1)}$) is automatically fixed
through the relation (\ref{friedmodif}). In order to obtain $H(z)$
we solve numerically the equation (\ref{friedmodif2}).

Since the most solid evidence for the acceleration of the universe comes
from measurements of luminosity distances for type Ia supernovae, we will
use the most recent compilation of 307 SNe Ia events (the \emph{Union}
sample) \cite{kowalski}, to put constraints in the $n-\Omega_m$ plane. The
predicted distance modulus for a supernova at redshift $z$, for a given set
of parameters \textbf{P}=($n$, $\Omega_m$), is
\begin{equation}  \label{mu}
\mu(z\mid\mathbf{P})=m-M=5\:log(d_L)+25
\end{equation}
where $m$ and $M$ are the apparent and absolute magnitudes respectively, and
$d_L$ stands for the luminosity distance (in units of megaparsecs),
\begin{equation}  \label{dl}
d_L(z;\mathbf{P})=(1+z)\:\int_{0}^{z}\frac{dz^{\prime}}{H(z^{\prime},\mathbf{%
P})}
\end{equation}
where $H(z;\mathbf{P})$ is given by the numerical solution of
(\ref{friedmodif2}). We use a $\chi^2$ statistic to find the
best-fit for a set of parameters \textbf{P} (marginalizing over
$H_o$),
\begin{equation}  \label{chi}
\chi^2_{SNe}=\sum_{i=1}^{N=307}\frac{[\mu_i(z\mid\mathbf{P}%
)-\mu_i^{obs}(z)]^2}{\sigma_i^2}
\end{equation}
where $\mu_i(z\mid\mathbf{P})$ is defined by (\ref{mu}), $\mu_i^{obs}$ and $%
\sigma_i$ are the distance modulus and its uncertainty for each
observed value \cite{kowalski}. As it is known, the measurements
of SNe Ia are not enough to constraint $\Omega_m$ thoroughly. To
perform the statistic we also consider, on one hand, the
information coming from the BAO (Baryon Acoustic Oscillation) peak
detected in the correlation function of luminous red galaxies
(LRG) in the Sloan Digital Sky Survey \cite{eisen}. The observed
scale of the peak effectively constraints the quantity (assumed a
$\Lambda CDM$ model),
\begin{equation}  \label{A035}
A_{0.35}=D_V(0.35)\:\frac{\sqrt{\Omega_m H_o^2}}{0.35}=0.469\pm0.017
\end{equation}
where $z=0.35$ is the typical redshift of the LRG and $D_V$ is
defined as
\begin{equation}  \label{DV}
D_V(z)=\Bigg[\frac{z}{H(z)}\:\Big(\int_{0}^{z}\frac{dz^{\prime}}{%
H(z^{\prime})}\Big)^{2}\Bigg]^{1/3}
\end{equation}

\begin{figure}[tbp]
\begin{center}
\includegraphics[width=8cm,angle=0]{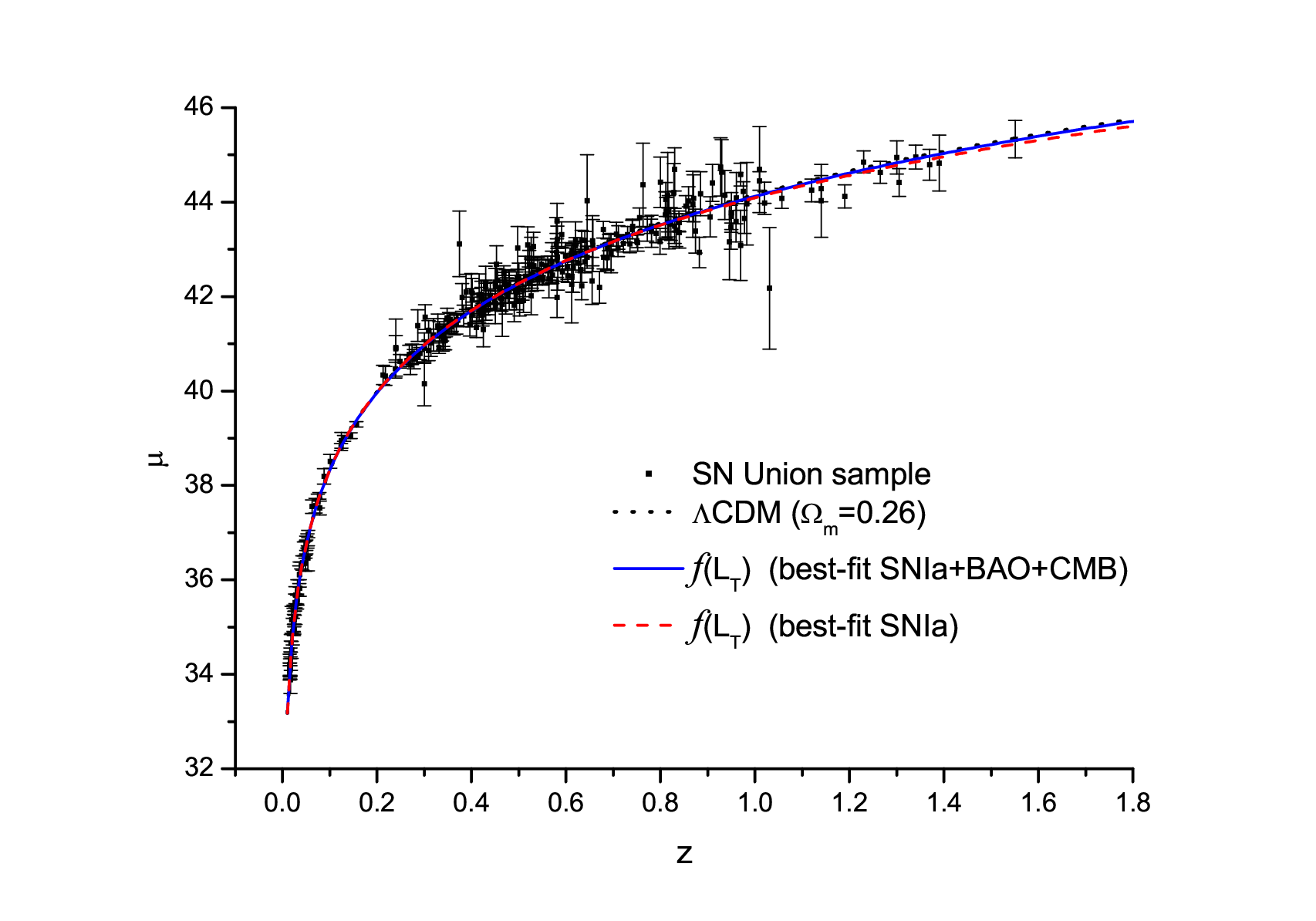}
\end{center}
\caption{Hubble diagram for 307 SNe Ia from the \emph{Union} sample
\protect\cite{kowalski}. The curves correspond to the concordance model $%
\Lambda CDM$ with $\Omega_\Lambda$=0.74 and $\Omega_m$=0.26
(dotted line), and our models with the values corresponding to the
best-fit $\Omega_m$=0.42 and $n$=1.30 (dashed line) and also with
those coming from the joint analysis of SNIa+BAO+CMB,
$\Omega_m$=0.27 and $n$=-0.10 (solid line).} \label{datasn}
\end{figure}

On the other hand, we have also included in the statistic the CMB
shift parameter, which relates the angular diameter distance to
the last scattering surface with the angular scale of the first
acoustic peak in the CMB power spectrum. In order to do this, we
have considered a radiation component $\Omega_r$=5 x $10^{-5}$.
The CMB shift parameter is given by \cite{komatsu},
\begin{equation}  \label{R1089}
R_{1089}=\sqrt{\Omega_m
H_o^2}\:\int_{0}^{1089}\frac{dz}{H(z)}=1.710\pm0.019
\end{equation}
We can use both parameters since our model presents matter domination at the
decoupling time. Figure \ref{datasn} shows the Hubble diagram for the 307
SNe Ia belonging to the \emph{Union} sample. The curves represent models
with values of $\Omega_m$ and $n$ obtained from minimizing $\chi^2$ using
only SNIa and SNIa+BAO+CMB as well. It is also shown as a reference the $%
\Lambda CDM$ model with $\Omega_m=0.26$. The obtained values for
the best-fit to the SNe Ia data only are $\Omega_m=0.42$ and
$n=1.30$ with the reduced
$\chi_\nu^2\equiv\chi_{min}^2/\nu\simeq1.02$ (or equivalently,
$\Delta\chi_{min}^2=-1.1$) where $\nu$ is the number of degrees of
freedom.

\begin{figure}[tbp]
\begin{center}
\includegraphics[width=8cm,angle=0]{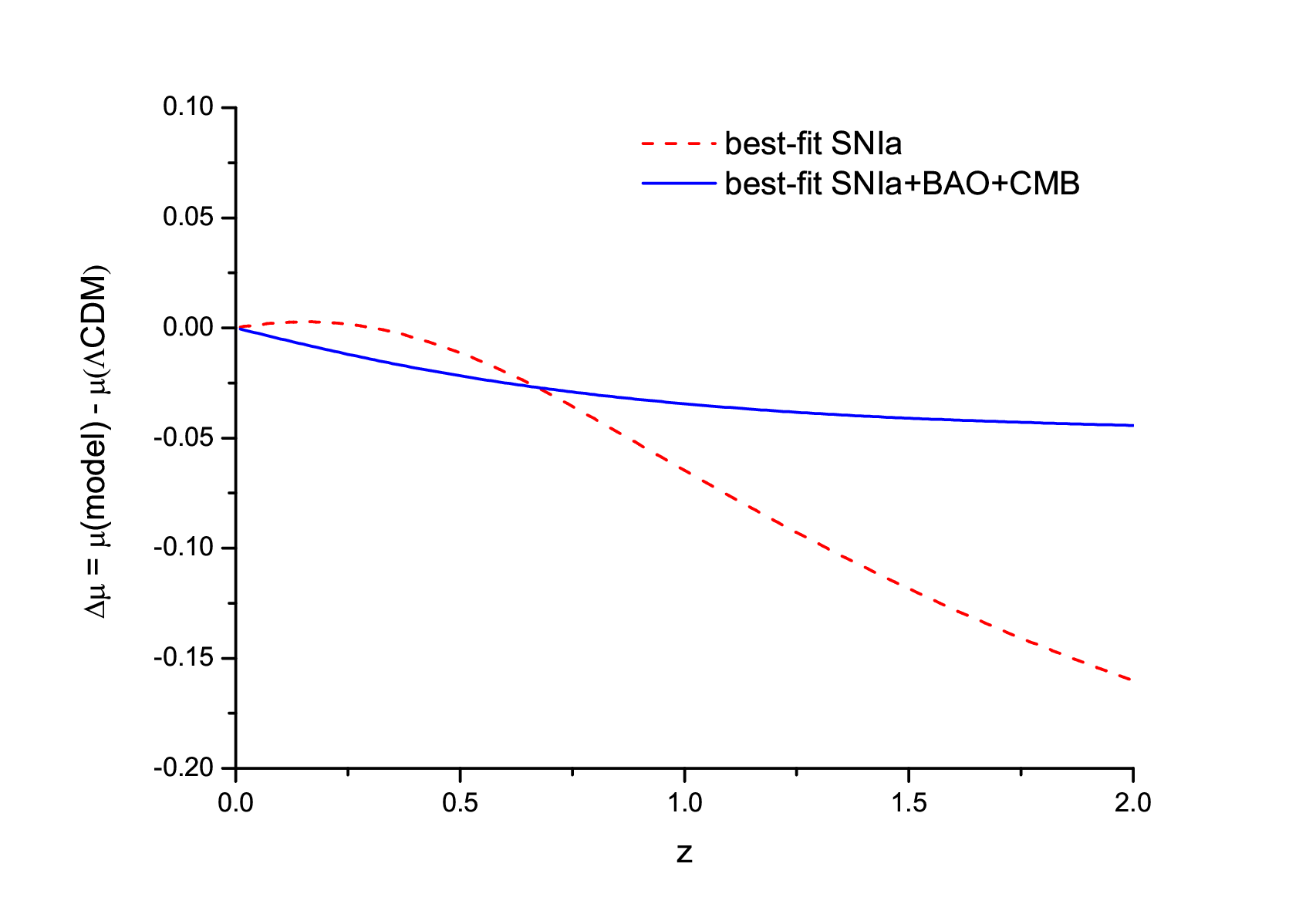}
\end{center}
\caption{Distance modulus residual from the $\Lambda CDM$ model
for the same values from Figure \ref{datasn}.} \label{deltamu}
\end{figure}

In Figure \ref{deltamu} we plot the distance modulus residual
($\Delta\mu$) from the $\Lambda CDM$ model to better appreciate
the discrepancies between our model and $\Lambda CDM$.

\begin{figure}[tbp]
\begin{center}
\includegraphics[width=8cm,angle=0]{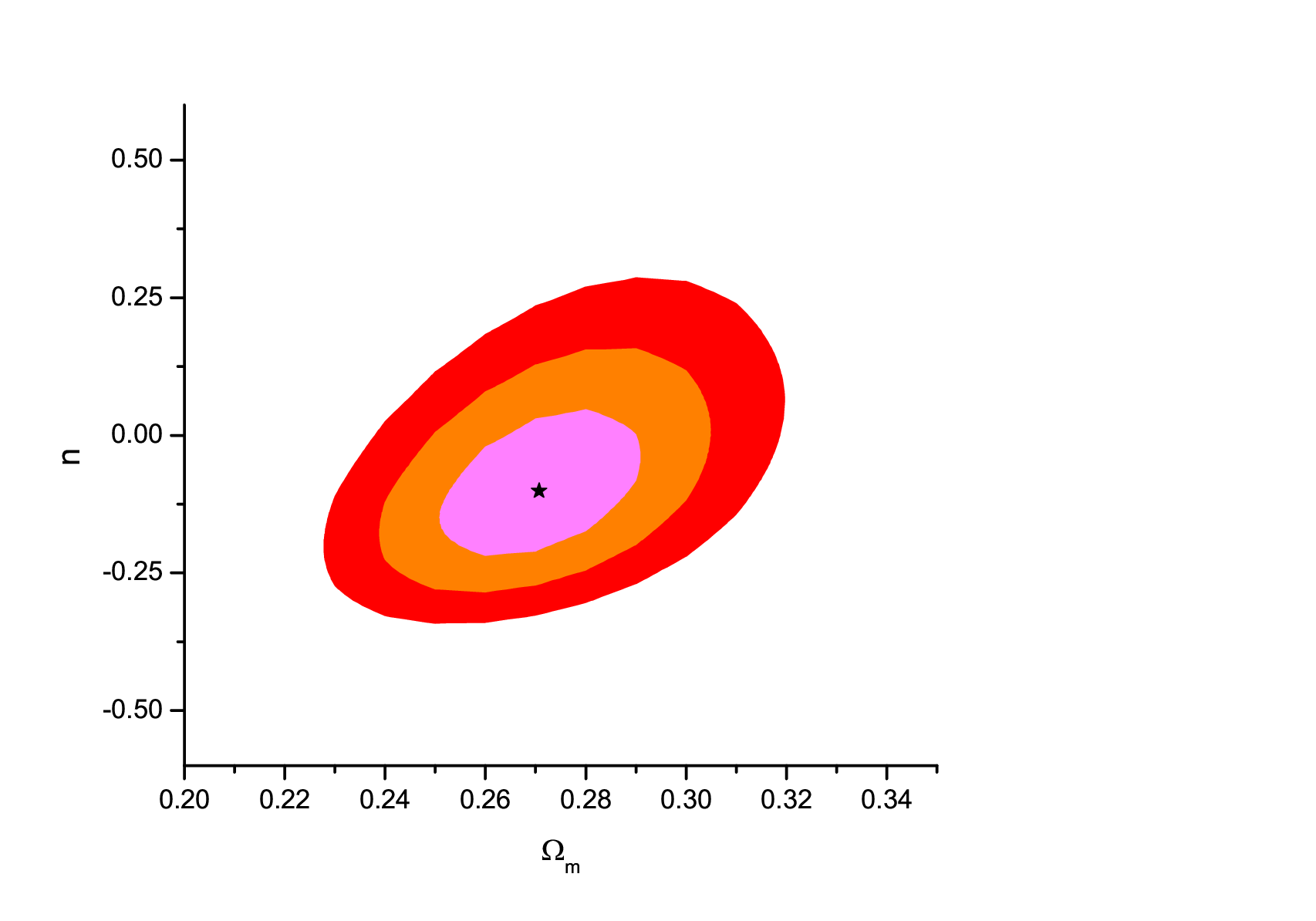}
\end{center}
\caption{Confidence intervals at 68.3\%, 95.4\% y 99.7\% in the
$n-\Omega_m$ plane coming from combining SNe Ia, BAO and CMB data.
The best-fit to this joint analysis is reached with the values
$n=-0.10$ and $\Omega_m=0.27$.} \label{suma}
\end{figure}

Figure \ref{suma} shows the confidence intervals at 68.3\%, 95.4\%
and 99.7\% for the joint probability of the parameters $n$ and
$\Omega_m$, having combined the SNe Ia data with BAO and CMB
parameters. This analysis
yields that the best-fit to all data is achieved with $n=-0.10$ and $%
\Omega_m=0.27$ (with a $\chi_{min}^2/\nu\simeq1.01,
\Delta\chi_{min}^2=-1.2$) and also the values of the parameters
lie in the ranges (at 68.4\% c.l.): $n\in[-0.23, 0.03];
\:\Omega_m\in$[0.25, 0.29].

For our model we have analyzed as well the total and effective
equations of state as a
function of $z$. From (\ref{friedmodif}) and (\ref{ec1-1}) along with (\ref%
{fT}), one can define a torsion contribution to the density and pressure as
\begin{eqnarray}
\rho_{T}&=&\frac{3}{8\pi G}\:\frac{(2n+1)\:\alpha}{6^{n+1}H^{2n}}  \nonumber
\\
p_T&=&\frac{\alpha}{8\pi G}\:\Big[(6H^2)^{-(n+1)}\dot{H}[4n(n+1)-2n]+
\nonumber \\
&-&6n(6H^2)^{-(n+1)}H^2 -\frac{(6H^2)^{-n}}{2}\Big]  \label{roypt}
\end{eqnarray}
to rewrite the dynamical equations as
\begin{eqnarray}  \label{friedc}
H^2&=&\frac{8\pi G}{3}(\rho+\rho_T) \\
\frac{\ddot{a}}{a}&=&-\frac{8\pi G}{6}[\rho+\rho_T+3(p+p_T)]  \label{1-1c}
\end{eqnarray}
Then, by using (\ref{friedc}) and (\ref{1-1c}) the total and
effective equations of state are written as,
\begin{equation}  \label{ecestado}
w_{tot}\equiv\frac{p+p_T}{\rho+\rho_T}=-1+\frac{2(1+z)}{3
H}\:\frac{dH}{dz}
\end{equation}
\begin{equation}  \label{ecestadoef}
w_{eff}=\frac{p_T}{\rho_T}
\end{equation}

\begin{figure}[tbp]
\begin{center}
\includegraphics[width=8cm,angle=0]{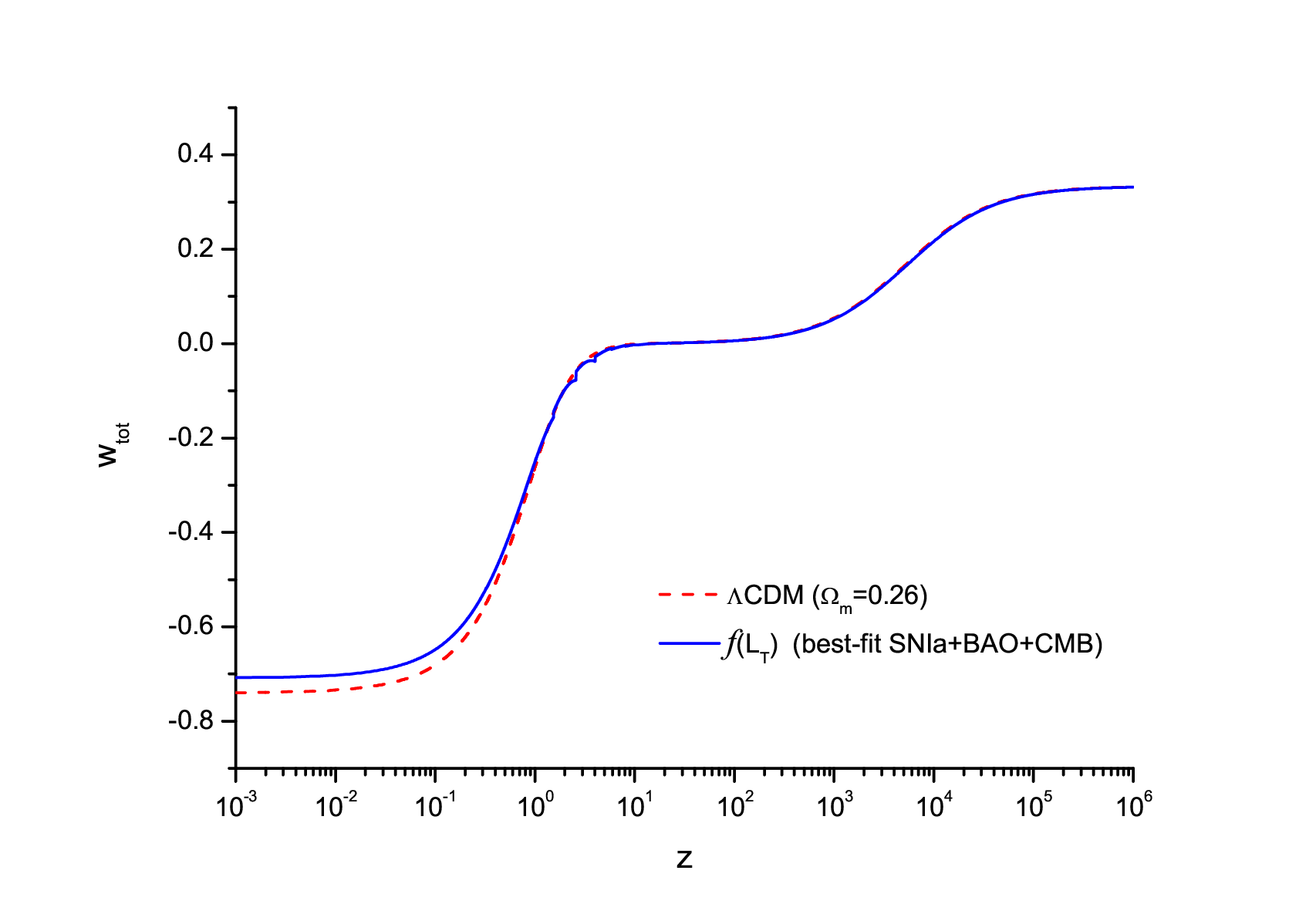}
\end{center}
\caption{The curves correspond to the total equation of state as a
function of $z$ expected for the standard concordance model
$\Lambda CDM$ with $\Omega_\Lambda$=0.74 and $\Omega_m$=0.26
(dashed line), and for our model (solid line) with the values of
the best-fit coming from SNIa+BAO+CMB, $\Omega_m$=0.27 and
$n$=-0.10. Three cosmological phases observed.} \label{wtot}
\end{figure}

\begin{figure}[tbp]
\begin{center}
\includegraphics[width=8cm,angle=0]{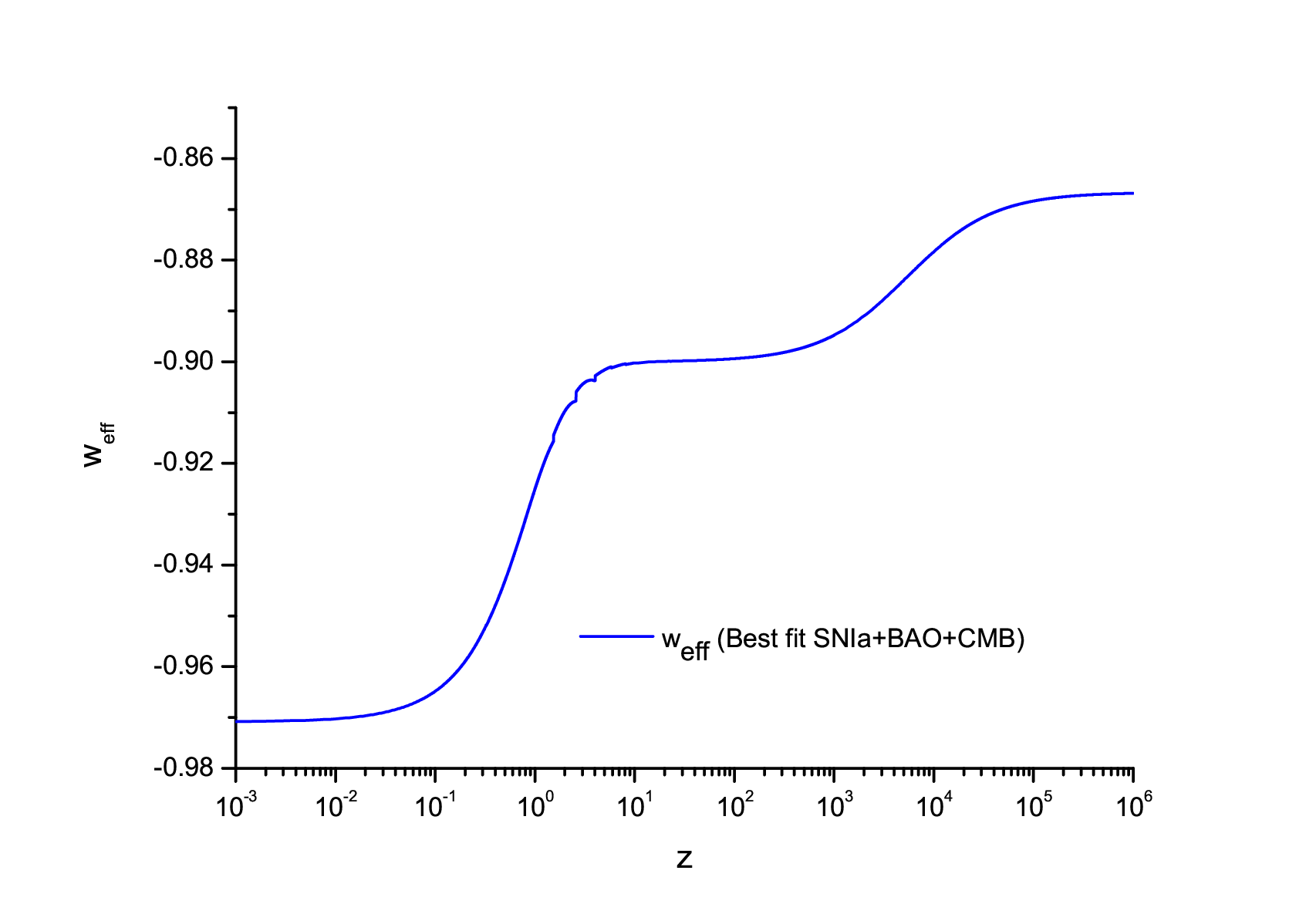}
\end{center}
\caption{Effective equation of state as a function of $z$ for our
model with the values of the best-fit coming from SNIa+BAO+CMB,
$\Omega_m$=0.27 and $n$=-0.10.} \label{weff}
\end{figure}

Figure \ref{wtot} shows the evolution of the total equation of state $%
w_{tot}$ as a function of $z$ for our model with the values of the
best-fit. There can be observed the last three phases of the
evolution of the universe: radiation dominated ($w=1/3$), matter
dominated ($w=0$) and late acceleration ($w\simeq-1$). Figure
\ref{weff} shows the effective equation of state coming from the
dark torsion contribution. Finally, analyzing (\ref{1-1c}) we find
that our model with $n=-0.10$ predicts that the transition from
deceleration to acceleration occurs at $z_{acc}\simeq0.74$ in good
agreement with recent works \cite{melch07}.

An interesting point to be highlighted is that equation
(\ref{friedmodif}) reveals that a value of $n>0$, as the one
obtained by considering only SNIa data, implies that the effective
dark torsion is of the phantom type \cite{cal02}. That is, since
$H$ decreases toward the present time, the dark torsion density
increases instead of diluting with expansion ($w_{eff}<-1$).
However, when combining the complete data with SNIa+BAO+CMB we can
see from Fig. \ref{suma} that it is slightly favored ($1\sigma$
c.l.) a model with $n\leqslant0$.

\section{Conclusions}

A theory $f(L_T)$ based on a modification of the teleparallel equivalent of
General Relativity (TEGR) -where torsion is the geometric object describing
gravity instead of curvature and its equations are always of 2nd order- is
remarkably simpler than $f(R)$ theories. We have tested the theory $%
f(L_T)=L_T-\alpha(-L_T)^{-n}$ with the aim of reproducing the recently
detected acceleration of the universe without resorting to dark energy. We
have here performed observational viability tests for this theory by using
the most recent SN Ia data, and combined them with the information coming
from BAO peak and CMB shift parameter in order to find constraints in the $%
n-\Omega_m$ plane. At 68.3\% c.l. we found that the values lie in
the ranges $n\in[-0.23, 0.03]$ and $\Omega_m\in[0.25, 0.29]$. The
values for $\Omega_m$ are consistent with recent estimations
obtained by other authors (see, e.g., \cite{spergel}). The model
with the best-fit values minimizing the $\chi^2$ that combines
SNIa+BAO+CMB data ($n=-0.10$ and $\Omega_m=0.27$) exhibits the
last three phases of cosmological evolution: radiation era, matter
era and late acceleration; this last stage having started at
$z_{acc}\simeq0.74$.

\bigskip

\bigskip

\acknowledgments{G.R.B. is supported by a CONICET graduate
scholarship. G.R.B thanks Franco Fiorini for his encouraging at
the early stages of the work. Both authors are grateful to Diego
Travieso, Gast\'{o}n Giribet and Susana Landau for helpful
discussions. This work was partially supported by CONICET.}

\end{document}